\documentclass{PoS}

\usepackage{dcolumn}
\usepackage{bm}
\usepackage{lscape}
\usepackage{amsmath,amssymb,graphicx}
\usepackage{epsfig}
\usepackage{pstricks}
\usepackage{multirow}
\usepackage{booktabs}
\usepackage{axodraw4j}
\usepackage{subfigure}
\usepackage{cite}
\usepackage{rotating}
\usepackage{float}
\usepackage{soul}
\usepackage{comment}
\usepackage{textcomp}
\usepackage{afterpage}

\usepackage{url}
\usepackage{picins}
\usepackage{caption}

\usepackage{array}
\newcolumntype{L}[1]{>{\raggedright\let\newline\\\arraybackslash\hspace{0pt}}m{#1}}
\newcolumntype{C}[1]{>{\centering\let\newline\\\arraybackslash\hspace{0pt}}m{#1}}
\newcolumntype{R}[1]{>{\raggedleft\let\newline\\\arraybackslash\hspace{0pt}}m{#1}}

\RequirePackage{lineno}

\newcommand{\Z}{\mathbb Z_2}

\newcommand{\MET}{\textbf{$E_{\rm T}^{\rm miss}$}}

\newcommand{\compresslist}{ 
\setlength{\itemsep}{1pt}
\setlength{\parskip}{0pt}
\setlength{\parsep}{0pt}
}

\title{Large width effects in processes of production of extra quarks decaying to Dark Matter at the LHC}

\ShortTitle{Production of XQs decaying to DM beyond the NWA at the LHC}

\author{\speaker{Hugo Prager}\\ 
        University of Southampton, Rutherford Appleton Laboratory\\
        E-mail: \email{hugo.prager@soton.ac.uk}}

\author{Stefano Moretti\\
        University of Southampton, Rutherford Appleton Laboratory, CERN\\
        E-mail: \email{s.moretti@soton.ac.uk}}

\author{Dermot O'Brien\\
        University of Southampton, Rutherford Appleton Laboratory\\
        E-mail: \email{doob1e14@soton.ac.uk}}

\author{Luca Panizzi\\
        Universit\`a di Genova and INFN, University of Southampton, Rutherford Appleton Laboratory\\
        E-mail: \email{luca.panizzi@ge.infn.it}}

\abstract{This paper explores the effects of finite width in processes of pair production of a heavy eXtra Quark (XQ) with charge 2/3 and its subsequent decay into a Dark Matter (DM) candidate -- either scalar or vector -- and Standard Model (SM) up-type quarks at the Large Hadron Collider (LHC). This dynamics has been ignored so far in standard experimental searches of heavy quarks decaying to DM and we assess herein the regions of validity of current approaches, based on the assumption that the XQ has a narrow width. Further, we discuss the configurations of masses, widths and couplings where the latter breaks down.}

\FullConference{XXV International Workshop on Deep-Inelastic Scattering and Related Subjects\\
		3-7 April 2017\\
		University of Birmingham, UK}

\begin{document}

\vspace{-5mm}

\section{Introduction}

\vspace{-2mm}

The question of the existence of new XQs and of the nature of DM is still opened after the SM. Here we consider a simplified model (representative of Universal Extra Dimensions (UED) models amongst others \cite{Antoniadis:1990ew,Servant:2002aq,Cacciapaglia:2009pa}) where the DM candidate is scalar or vector and produced through the decay of an XQ. We get signatures with $\MET$ similar to the supersymmetry (SUSY) ones which makes it possible to interpret SUSY results in the Narrow-Width Approximation (NWA) in terms of limits on XQs \cite{Cacciapaglia:2013wha,Kraml:2016eti}. Here we want to evaluate the effects of a \emph{large width} on the determination of the cross section and in the reinterpretation of bounds from experimental searches as we did in \cite{Moretti:2016gkr} for XQs decaying to SM particles. Note that the large width of the XQ implies its off-shellness which in turn requires, because of gauge invariance, the addition of the new graphs in the definition of the signal. \vspace{1.5mm}

\noindent \textbf{Lagrangian terms}
 
We use the following Lagrangian which describes the interaction between a \textit{singlet} DM and the XQ for a coupling with first generation SM quarks\footnote{The interaction terms for a coupling with second and third generation quarks are similar.} \vspace{-2mm}
\begin{eqnarray}
L^S_1 &=& 
\left[
\lambda_{11}^u \bar{T} P_R \; u + 
\lambda_{11}^d \bar{B} P_R \; d +
\lambda_{21} \; \overline\Psi_{1/6} P_L {u \choose d} 
\right] 
S^0_{DM} , \notag
\label{eq:LagSingletDMS}
\\
L^V_1 &=& 
\left[
g_{11}^u \bar{T} \gamma_\mu P_R \; u + 
g_{11}^d \bar{B} \gamma_\mu P_R \; d + 
g_{21}  \; \overline\Psi_{1/6} \gamma_\mu P_L {u \choose d} 
\right] 
V^{0\mu}_{DM} , \notag 
\label{eq:LagSingletDMV}
\end{eqnarray}
where $T$, $B$, $\Psi_{1/6}=(T \ B)^T$ are respectively two singlets and a doublet XQs, while $S^0_{DM}$ and $V^0_{DM}$ are scalar and vector DM. All these new particles are odd under a $\Z$ symmetry which is needed to make the DM stable. In the following we will only focus on a \emph{vector-like $T$} (a heavy top-quark partner) part of a doublet, since we checked that we obtain similar results for a singlet and also for chiral quarks.  \vspace{1.5mm}

\noindent \textbf{Observables and conventions}

We consider two different processes leading to the same final state $DM \; DM \ q \; \bar{q}$: \vspace{-2mm}
\begin{itemize} \compresslist
\item \textit{QCD pair production and on-shell decay}: this is the one considered in the experimental searches, using the NWA we have $\sigma_X (M_Q) \equiv \sigma_{2 \to 2}^{QCD}~BR(Q)~BR(\bar Q)$,
\item \textit{Full signal}: $\sigma_S (M_Q, M_{\rm DM}, \Gamma_Q)$ where all the topologies containing at least one XQ are taken into account, including some which are missing in the NWA,
\end{itemize}\vspace{-2mm}
and we study the ratio $(\sigma_S - \sigma_X)/\sigma_X$ which measures how much the full signal differs from the NWA one. \vspace{1.5mm}

\noindent \textbf{Analysis tools and experimental searches} 

To study the ratio of cross sections $(\sigma_S - \sigma_X)/\sigma_X$ as well the influence the width of the XQ on the mass bounds we consider an XQ top-partner $T$ and scan over the free parameters $M_T$, $M_{\rm DM}$ and $\Gamma_T$. We analyse in detail scenarios where the DM state has masses $M_{\rm DM}$ = 10 GeV and 500 GeV\footnote{The case of a DM state of mass 1 TeV has also been studied in the full paper \cite{Moretti:2017qby} but will not be presented here.} and with an XQ of mass $M_T > M_{\rm DM} + m_q$, with $q \in \{u,c,t\}$ (such that its on-shell decay is kinematically allowed) up to $M_T^{\rm max}$ = 2500 GeV, which is the maximal value of a $T$ mass so that it can be produced by the LHC@13TeV with 100/fb of integrated luminosity. We also consider values of the $T$ width from $\Gamma_T / M_T$ $\simeq$ 0\% (NWA) to 40\%.

Our numerical results at partonic level are obtained using {\sc MadGraph5} \cite{Alwall:2011uj,Alwall:2014hca} and a model we implemented in {\sc Feynrules} \cite{Alloul:2013bka} to obtain the UFO interface format. For the Monte Carlo simulation we use the PDF set {\sc cteq6l1}~\cite{Pumplin:2002vw}. Events are then passed to {\sc Pythia}\,8~\cite{Sjostrand:2007gs,Sjostrand:2006za}, which takes care of the hadronisation and parton showering. To analyse and compare the effects of a set of 13 TeV analyses considering final states compatible with our scenarios, we employ {\sc CheckMATE 2}~\cite{Dercks:2016npn}, which uses the {\sc Delphes\,3}~\cite{deFavereau:2013fsa} framework for the emulation of detector effects. In our simulations we include all the ATLAS and CMS (carried out at 13 TeV) analyses available within the CheckMATE database, the most relevant ones in our case being ATLAS 1604.07773 \cite{Aaboud:2016tnv}, ATLAS 1605.03814 \cite{Aaboud:2016zdn}, ATLAS-CONF-2016-050 \cite{ATLAS-CONF-2016-050}. \vspace{-2mm}

\section{Extra $T$ quark interacting with DM and the SM top quark}
\label{sec:Tt} \vspace{-2mm}

The possible decay channels for a XQ coupling to third generation SM quarks are $t \bar t + S^0_{DM} S^0_{DM}$ and $t \bar t + V^0_{DM} V^0_{DM}$, {i.e.} $t \bar t + \MET$.

\piccaption{\label{fig:SXthird} Relative difference between the full signal and the QCD pair production cross sections $(\sigma_S - \sigma_X)/\sigma_X$ in the $(M_T, \Gamma_T / M_T)$ plane. Left: scalar DM, $M_{\rm DM}$ = 10 GeV; right: vector DM, $M_{\rm DM}$ = 500 GeV.}
\parpic{
\epsfig{file=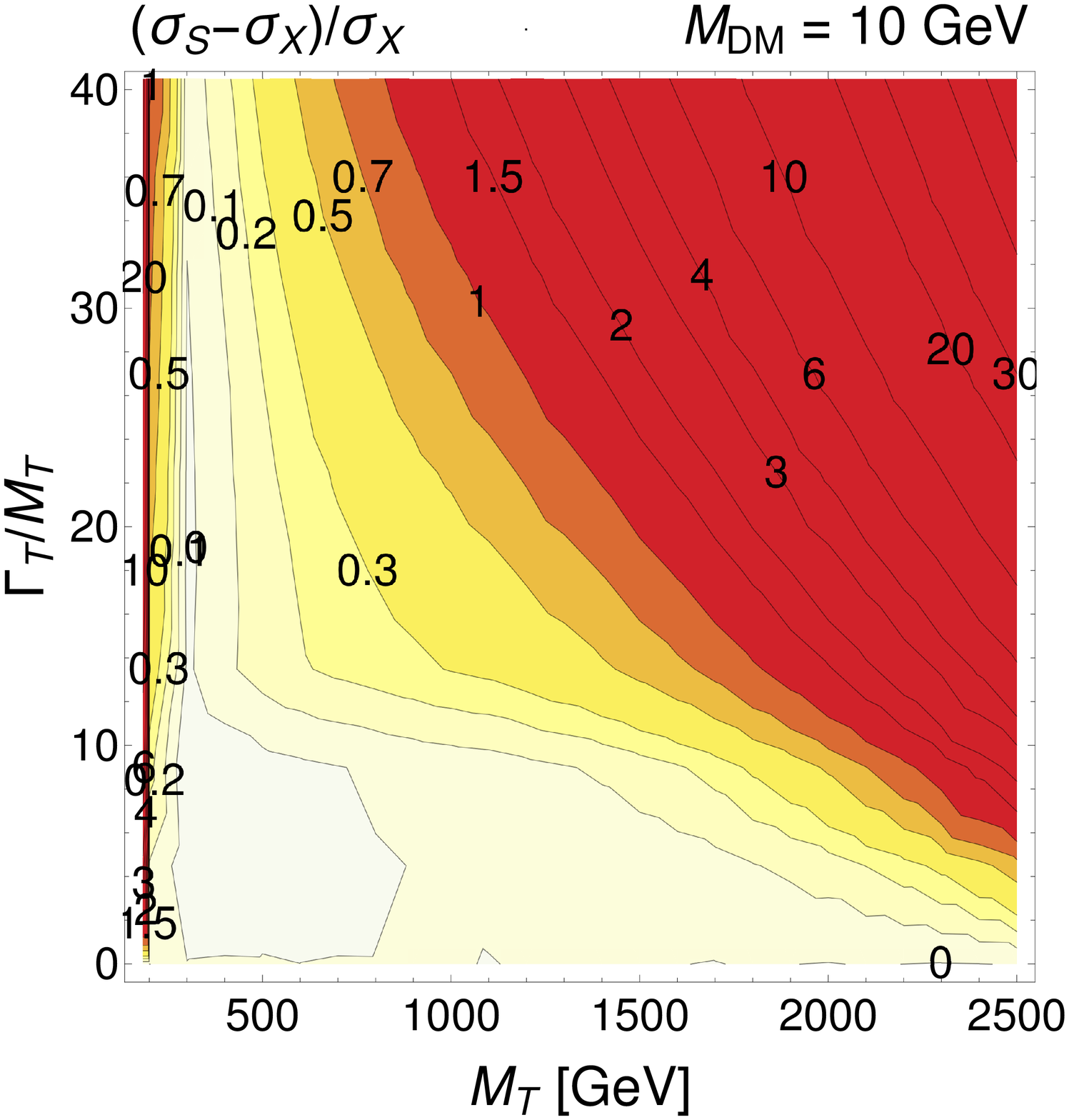, width=.26\textwidth} 
\epsfig{file=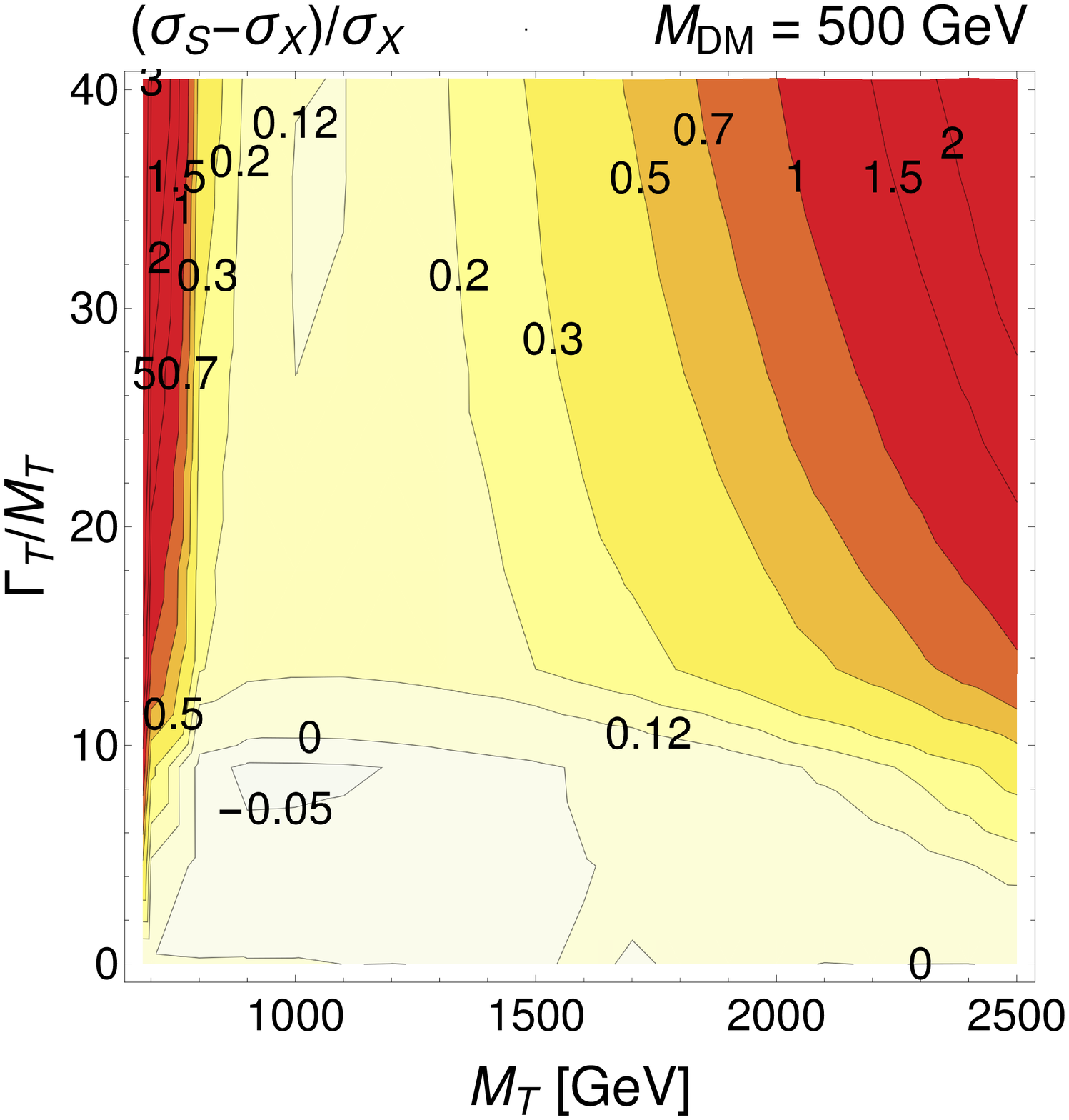, width=.26\textwidth}
}

\noindent \textbf{Large width effects at parton level}

\noindent We show in Fig. \ref{fig:SXthird} the relative difference between the full signal and the QCD pair production cross sections for an LHC energy of 13 TeV.
As expected we observe that any off-shell contribution is \emph{negligible} in the NWA and becomes more and more relevant when $\Gamma_T$ increases, especially when we are close to the kinematic threshold. The increase near the kinematics limit can be explained by a non-trivial combination of factors, the most relevant being the fact that a larger width opens a larger phase space for the decay of the $T$, which is more limited (in the NWA) as the gap between the masses decreases.  We also notice \emph{cancellations} that make $\sigma_X$ similar to $\sigma_S$ even for large values of the width, but it does not mean the NWA can be used in this region (we checked that the differential distributions are different).


\piccaption{\label{fig:Exclusion3}{\sc CheckMATE} results for DM particle of mass 10 GeV and 500 GeV. We show in black (grey) the exclusion line for the scalar (vector) DM scenario.}
\parpic[right]{
\epsfig{file=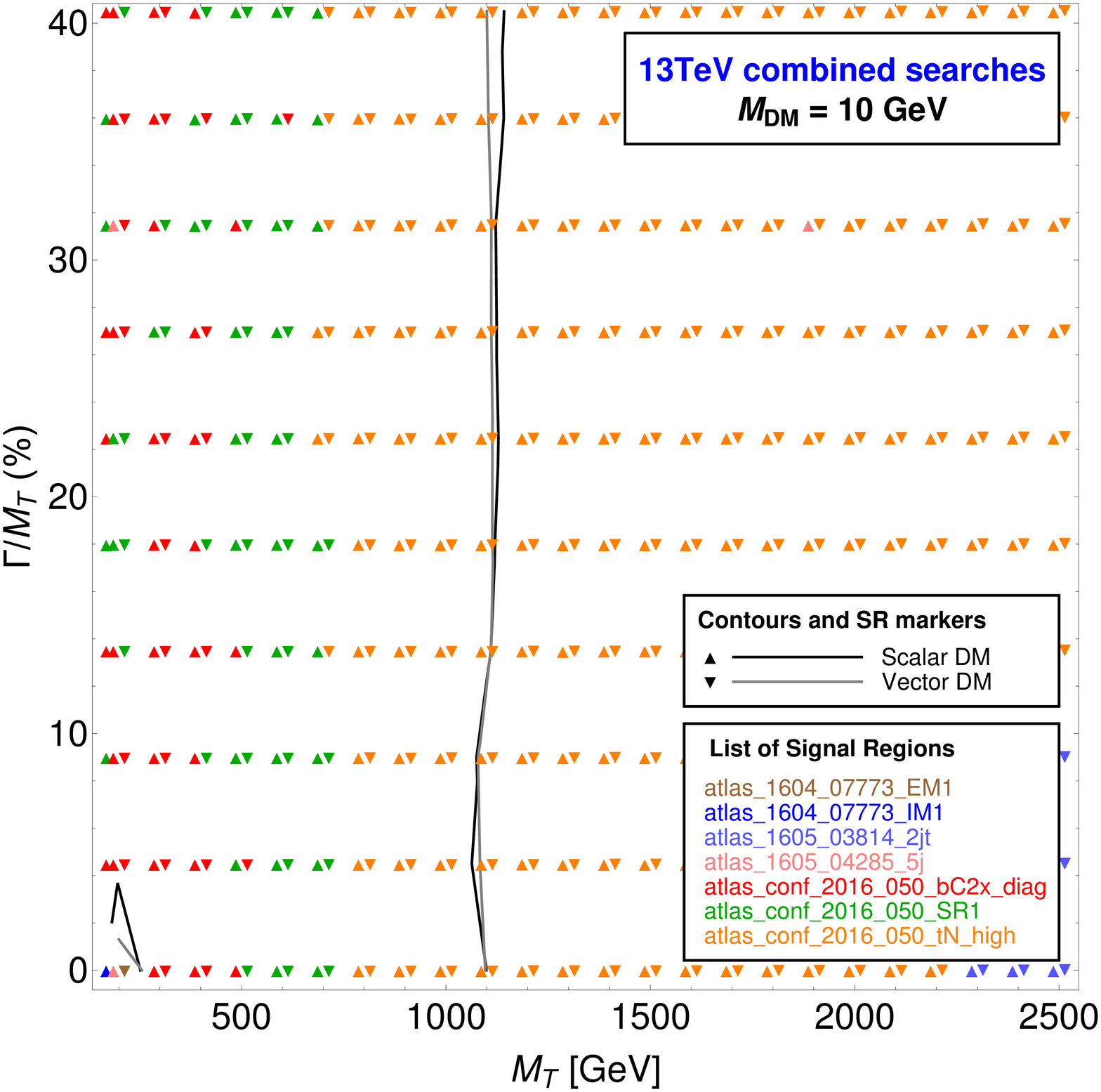, width=.26\textwidth} 
\epsfig{file=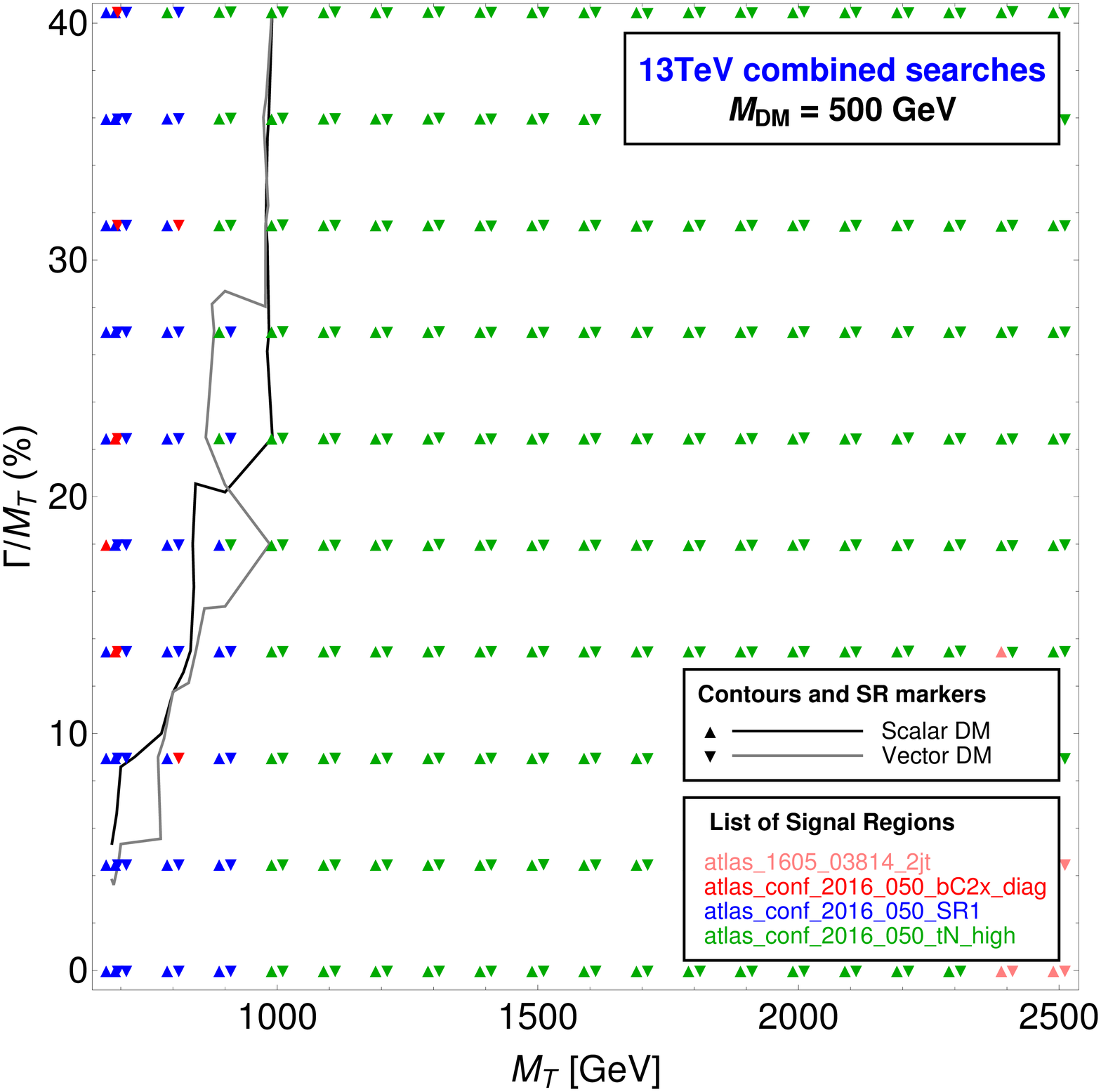, width=.26\textwidth} 
}

\noindent \textbf{Large width effects at detector level}

\noindent We show in Fig. \ref{fig:Exclusion3} the exclusion line for the scalar (vector) DM scenario in black (grey) and the best signal region. We notice that the bounds for scalar and vector DM very similar despite having different values of $\sigma_S$ in both cases, and that the width dependence is very small. The shape of the bounds is driven by the combination of cross section and selection efficiencies effect, the small width dependence being due to a combination of cuts on $p_T^{\rm jet}$ and $\MET$ leading to smaller efficiencies when $\Gamma_T$ increases.

\piccaption{\label{fig:Combined3} $\sigma_S$ (left) and efficiency (right) with the exclusion line for a scalar DM, $M_{\rm DM}$ = 10 GeV.}
\parpic[right]{
\epsfig{file=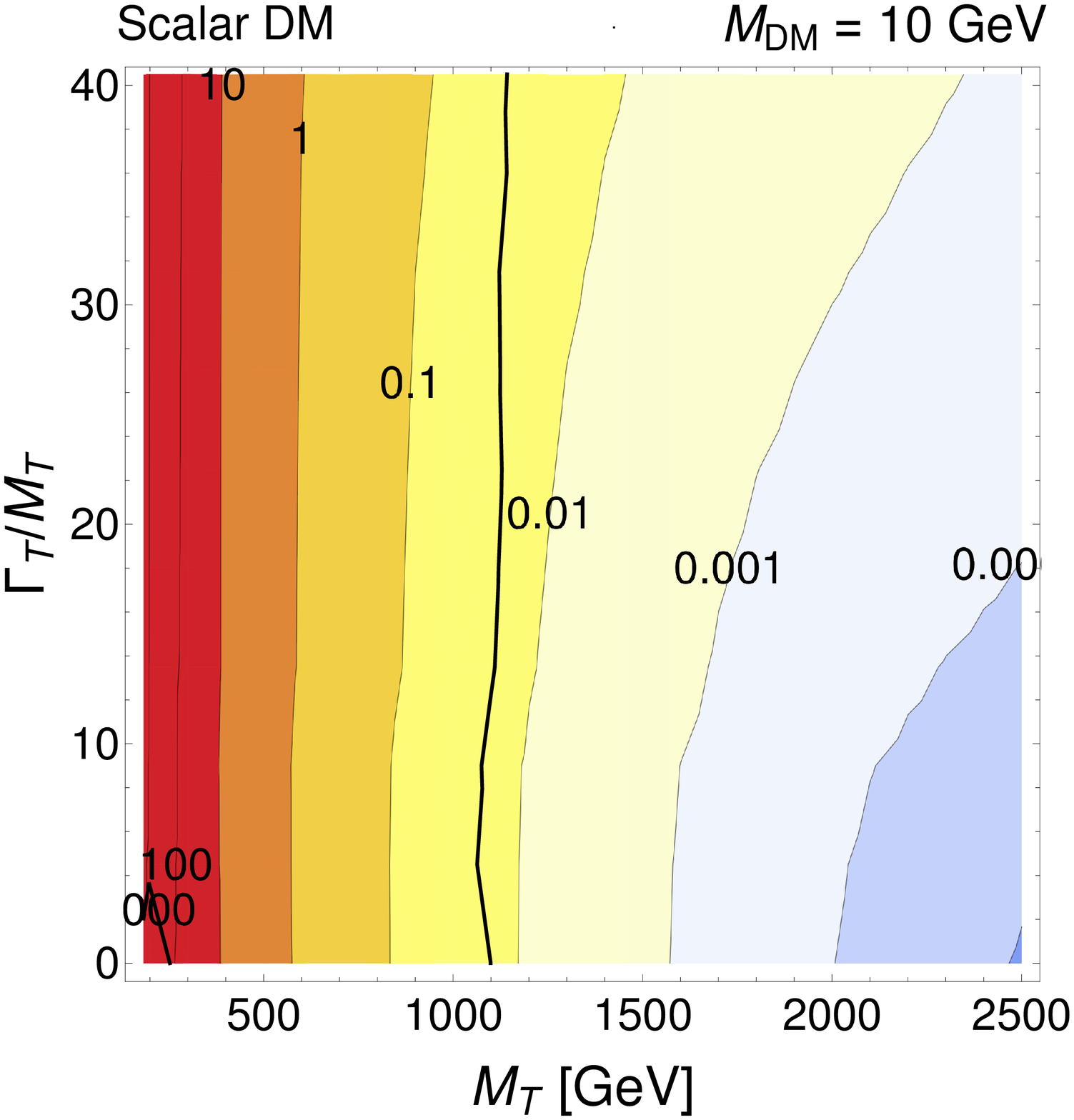, width=.26\textwidth} 
\epsfig{file=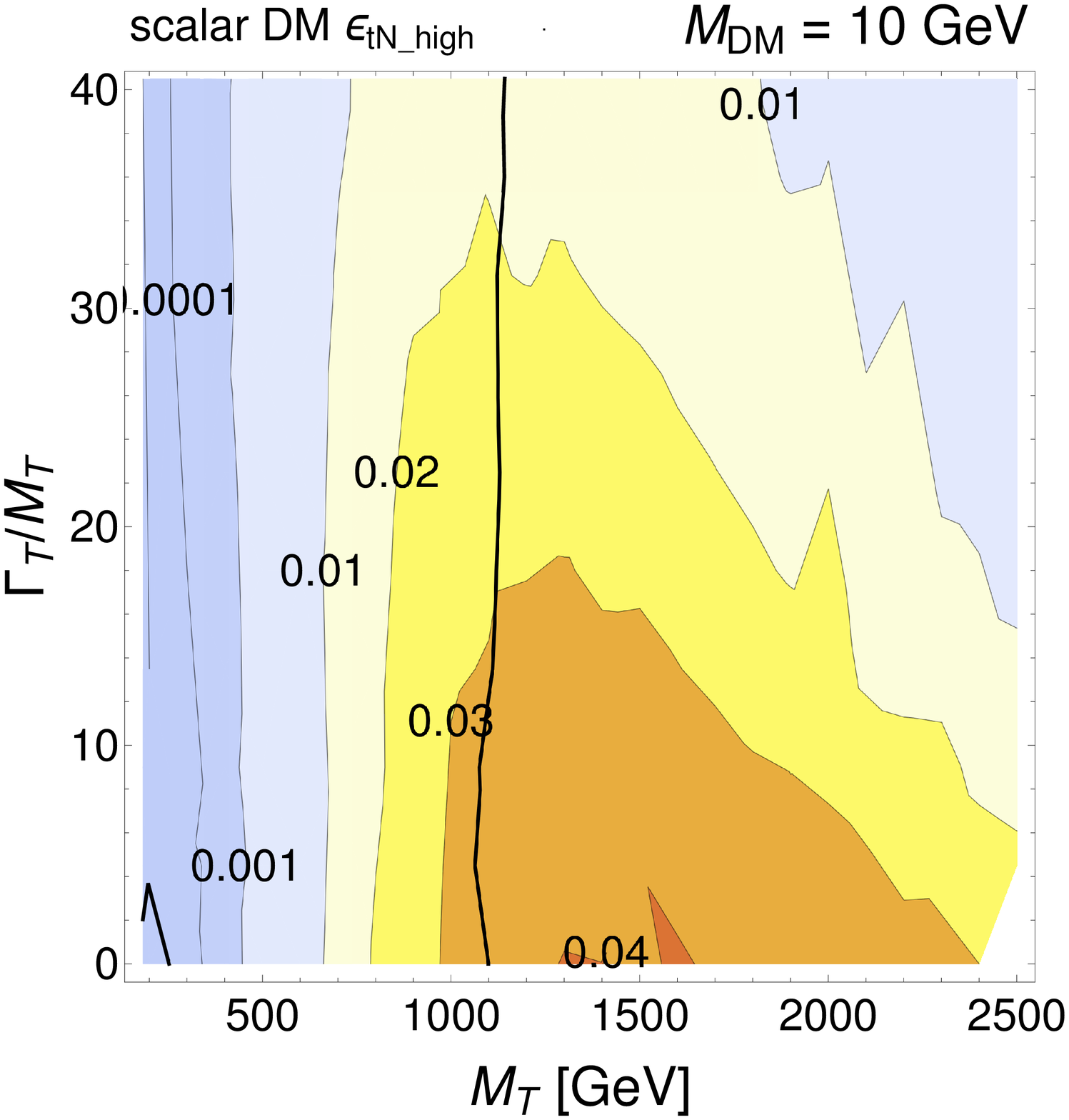, width=.26\textwidth} 
} \vspace{-5mm}

\noindent In Fig. \ref{fig:Combined3} the value of the signal cross section and efficiencies together with the exclusion line for a scalar DM of mass 10 GeV is shown. From these plots it is clear that the increase of cross section is compensated by the decrease of efficiency.
Therefore the hypothesis made for experimental searches are conservative since they \emph{never overestimate the bound}. This is due to the fact that $\sigma_S$ generally increases with the XQ width which leads to stronger bounds. 

\vspace{-4mm}

\section{Extra $T$ quark interacting with DM and the SM up quark}
\label{sec:Tu}

\vspace{-2mm}

\parpic[right]{
\begin{picture}(80,56)(0,-10)
\SetWidth{1}
\Line[arrow](0,0)(25,0)
\Text(-2,0)[rc]{\large $u$}
\Line[arrow](25,45)(0,45)
\Text(-2,45)[rc]{\large $\bar u$}
\Photon(20,0)(70,0){3}{9}
\Line[dash](25,0)(70,0)
\Text(72,0)[lc]{\large DM}
\SetColor{Red}\SetWidth{1.5}
\Line[arrow](25,0)(25,22.5)
\Text(21,12)[rc]{\large \Red{$T$}}
\Line[arrow](25,22.5)(25,45)
\Text(21,32)[rc]{\large \Red{$T$}}
\SetColor{Black}\SetWidth{1}
\Photon(25,45)(70,45){3}{9}
\Line[dash](25,45)(70,45)
\Text(72,45)[lc]{\large DM}
\Gluon(25,22.5)(50,22.5){3}{6}
\Line[arrow](50,22.5)(71,32)
\Text(72,32)[lc]{\large $u$}
\Line[arrow](70,10)(50,22.5)
\Text(72,10)[lc]{\large $\bar u$}
\end{picture}
} \vspace{-2mm}
We now study the case of XQs coupling to first generation SM quarks and a DM candidate. The possible final states are therefore $u \bar u + S^0_{DM} \; S^0_{DM}$ and $u \bar u +  V^0_{DM} \; V^0_{DM}$, {i.e.}  $j j + \MET$.
In this case new topologies containing infrared divergences (due to the gluon splitting) are present, such as the one showed on the right. These diagrams affect the signal but not the QCD pair production, therefore giving rise to a large increase of the ratio $(\sigma_S - \sigma_X) / \sigma_X$. For this reason we use logarithmic plots in the following.


\piccaption{\label{fig:SXfirst} Relative difference between the full signal and the QCD pair production cross sections $(\sigma_S - \sigma_X)/\sigma_X$ in the $(M_T, \Gamma_T / M_T)$ plane. Left: scalar DM, $M_{\rm DM}$ = 10 GeV; right: vector DM, $M_{\rm DM}$ = 500 GeV.}
\parpic{
\epsfig{file=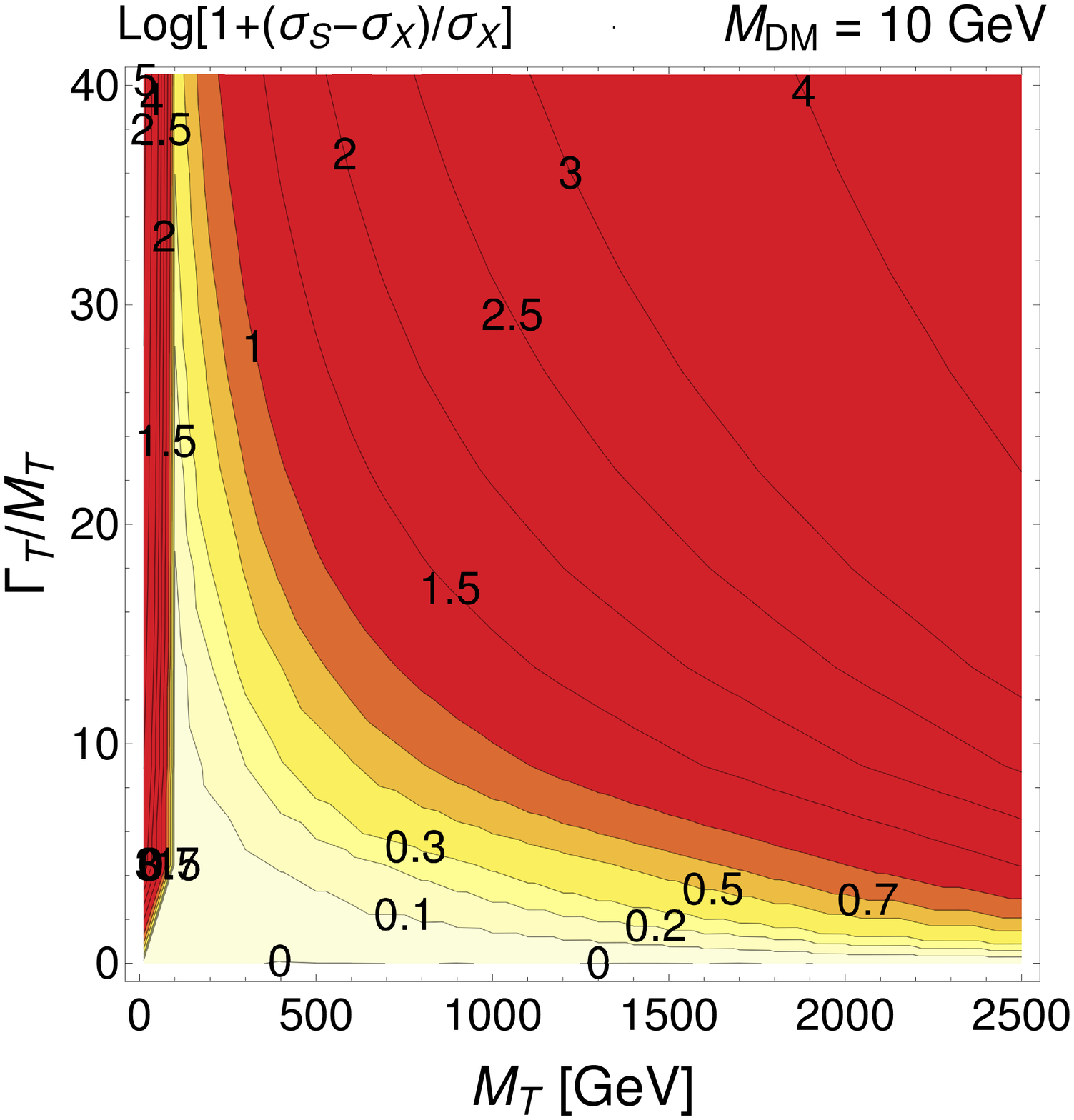, width=.26\textwidth} 
\epsfig{file=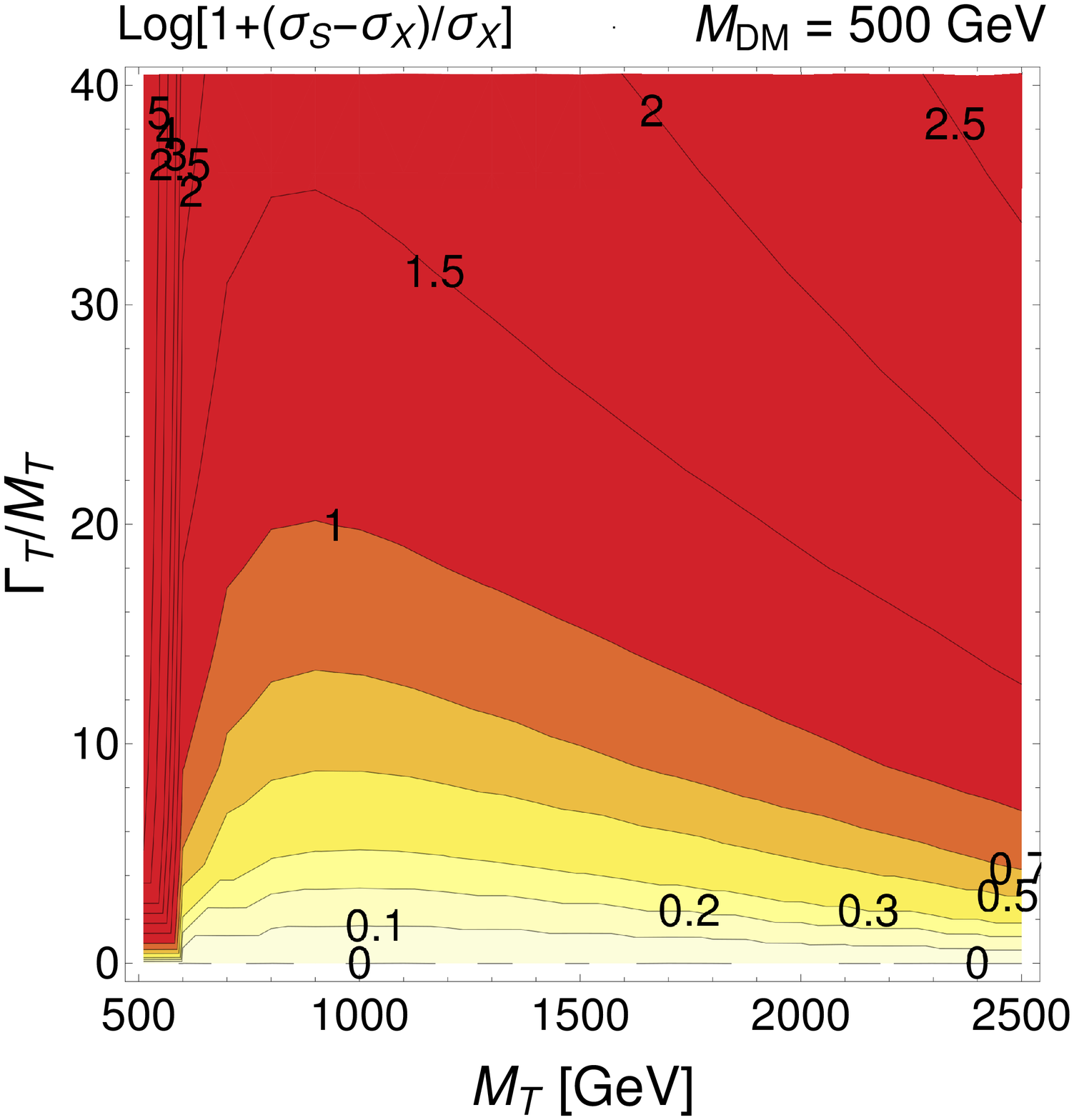, width=.26\textwidth} 
}

\noindent \textbf{Large width effects at parton level}

\noindent We see in Fig. \ref{fig:SXfirst} that the cross section ratio is much larger than for third generation coupling but is still zero in the NWA as expected. Also there is no cancellation of effects which makes $\sigma_S$ similar to $\sigma_X$ for large width, we only notice a decrease of the ratio in regions that are very similar to the ones observed for third generation. This means that the opposite effects we observed previously can only lower the signal but not cancel the additional diagrams.


\piccaption{\label{fig:Exclusion1}{\sc CheckMATE} results for DM particle of mass 10 GeV and 500 GeV. We show in black (grey) the exclusion line for the scalar (vector) DM scenario.}
\parpic[right]{
\epsfig{file=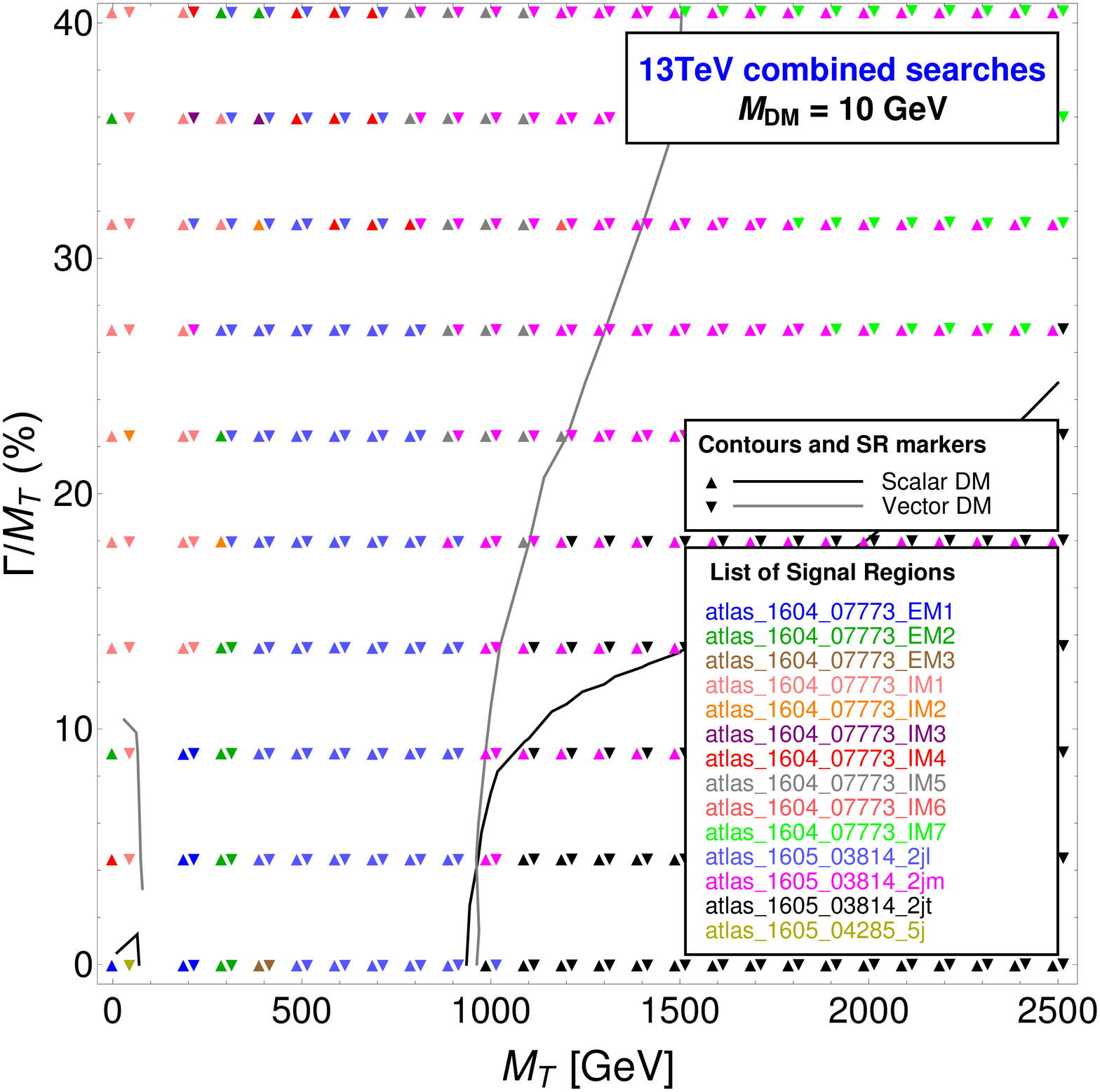, width=.26\textwidth} 
\epsfig{file=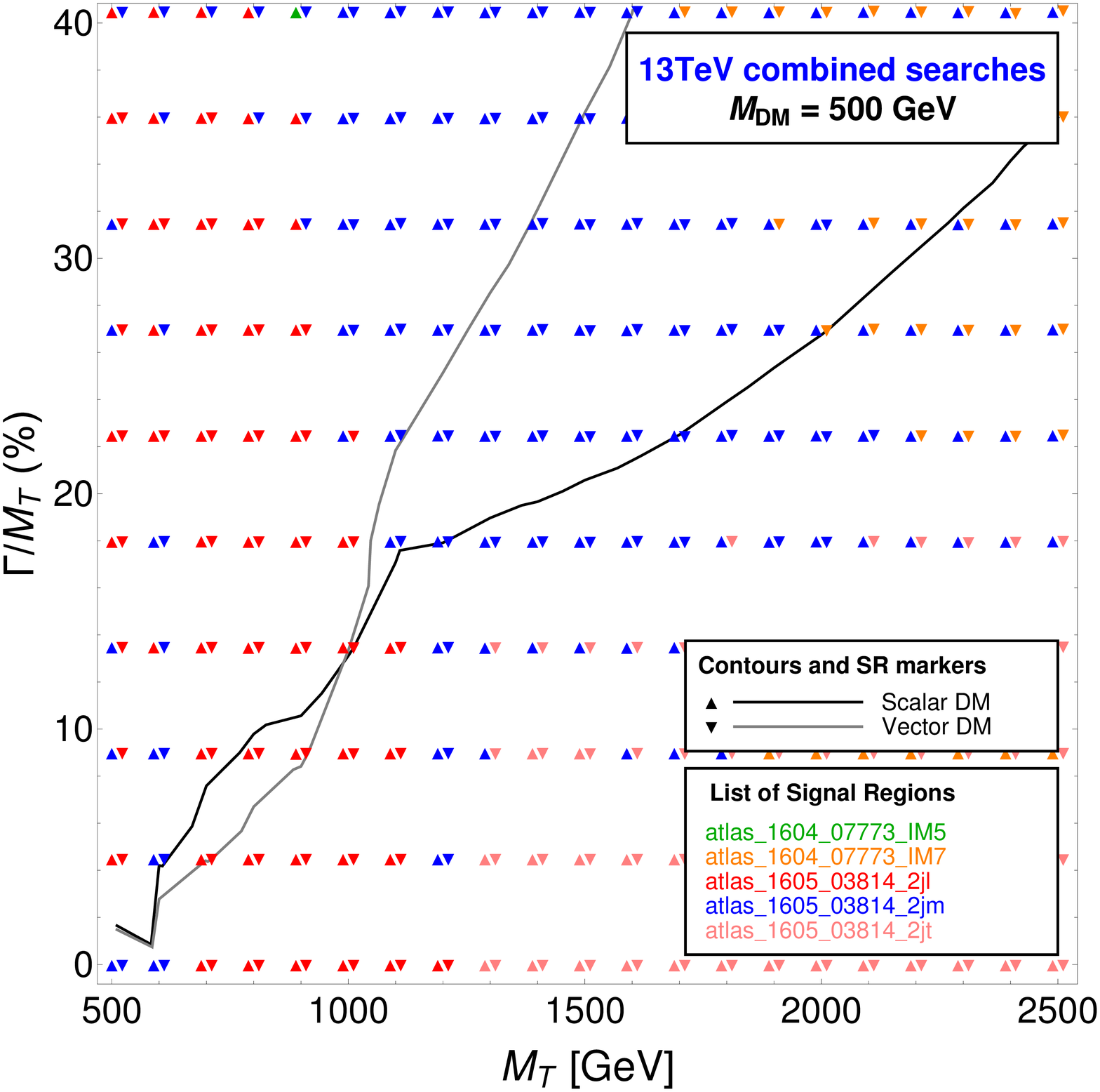, width=.26\textwidth} 
}

\noindent \textbf{Large width effects at detector level}

\noindent We show in Fig. \ref{fig:Exclusion1} the bounds for scalar (black line) and vector DM scenarios (grey line) together with the best signal region shown with a colour code. We observe an important width dependence of the bounds and it is possible to distinguish scalar and vector DM. The bounds are yet still similar for scalar and vector DM in the NWA because $\sigma_S (M_T, NWA) = \sigma_X (M_T, NWA) = \sigma_X (M_T)$ leading to the same exclusion. \\

\hspace{-4mm}\begin{minipage}{0.53\textwidth}
\includegraphics[width=.49\textwidth]{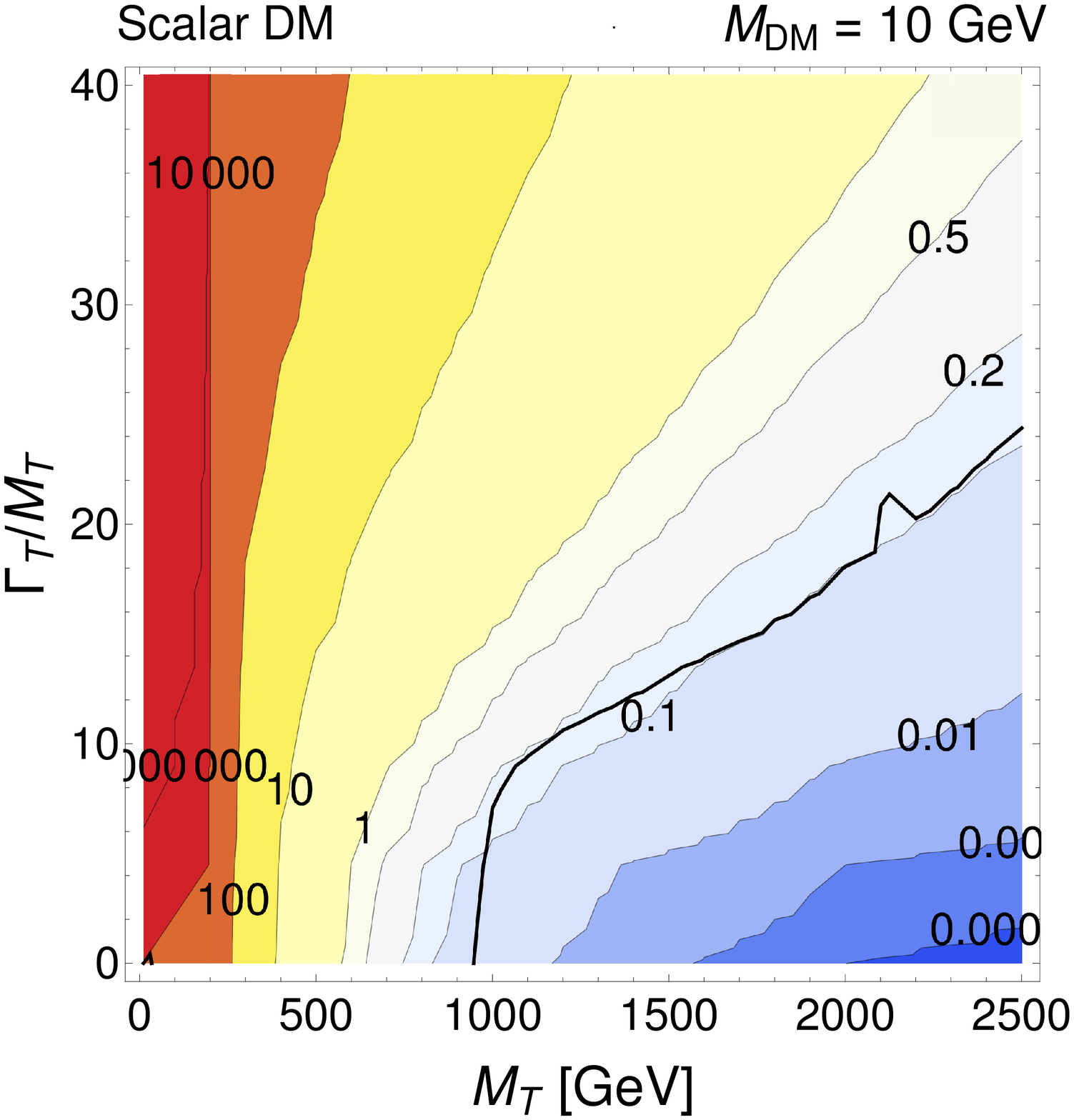}
\includegraphics[width=.49\textwidth]{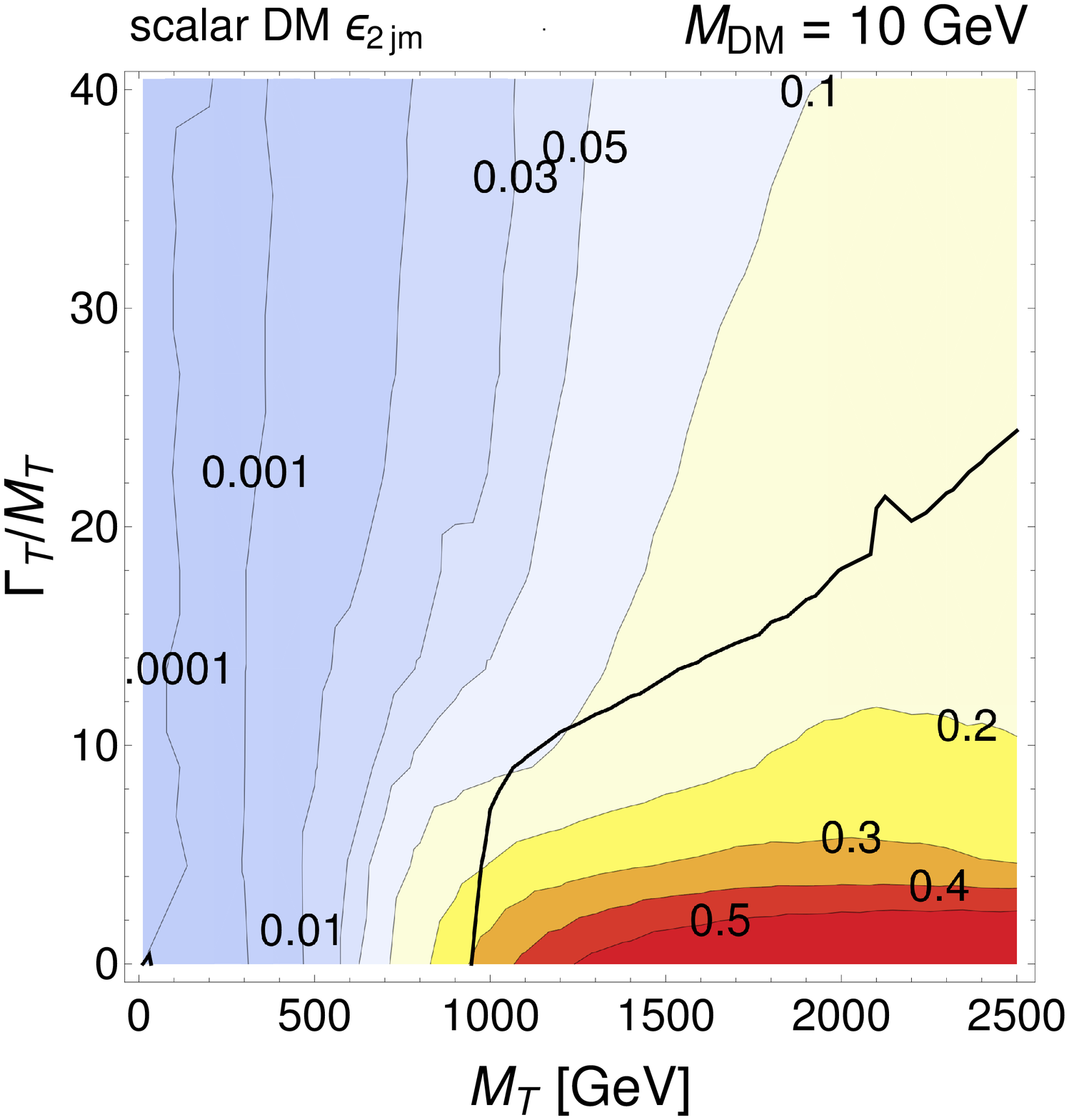} \\
\includegraphics[width=.49\textwidth]{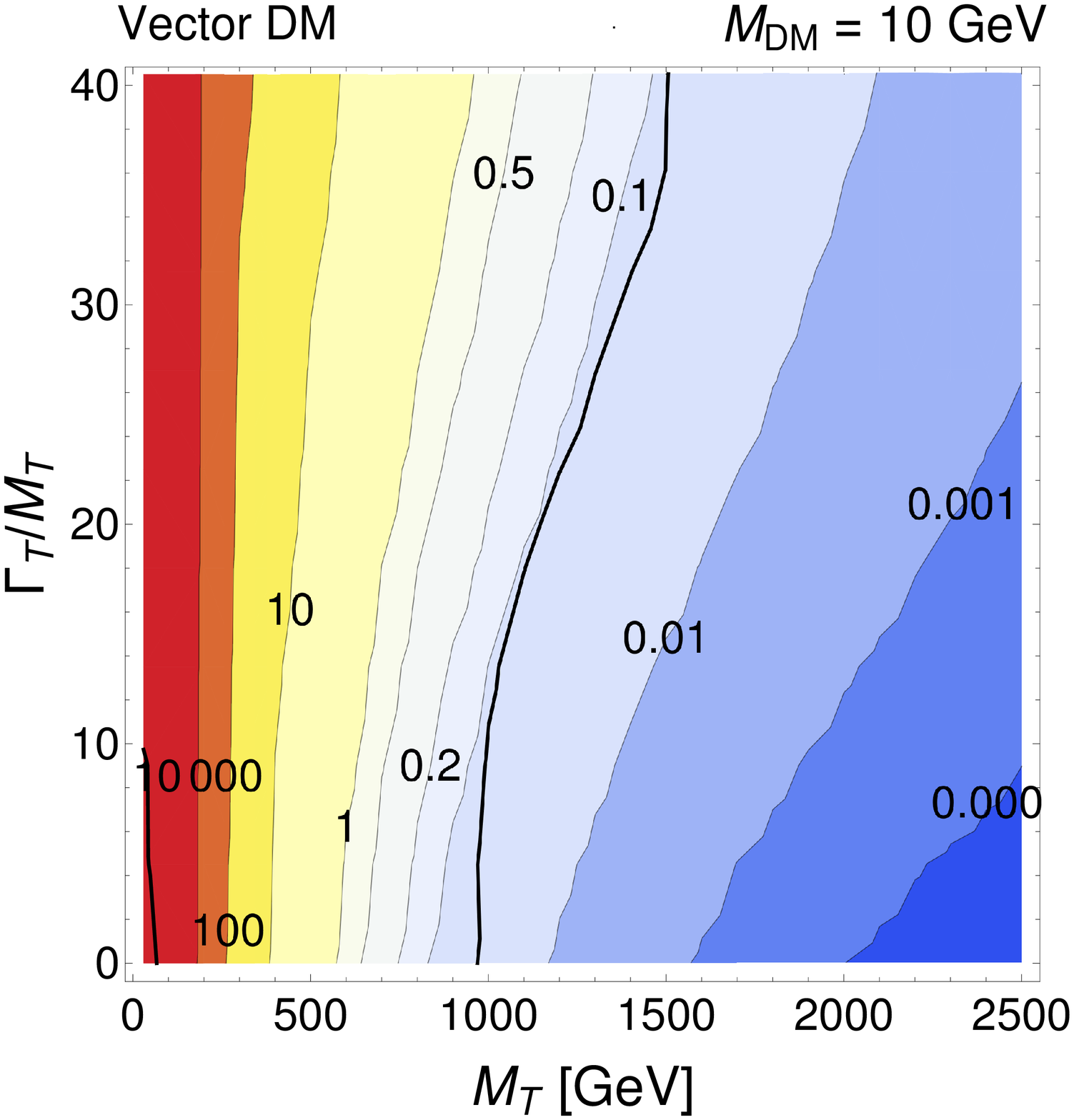}
\includegraphics[width=.49\textwidth]{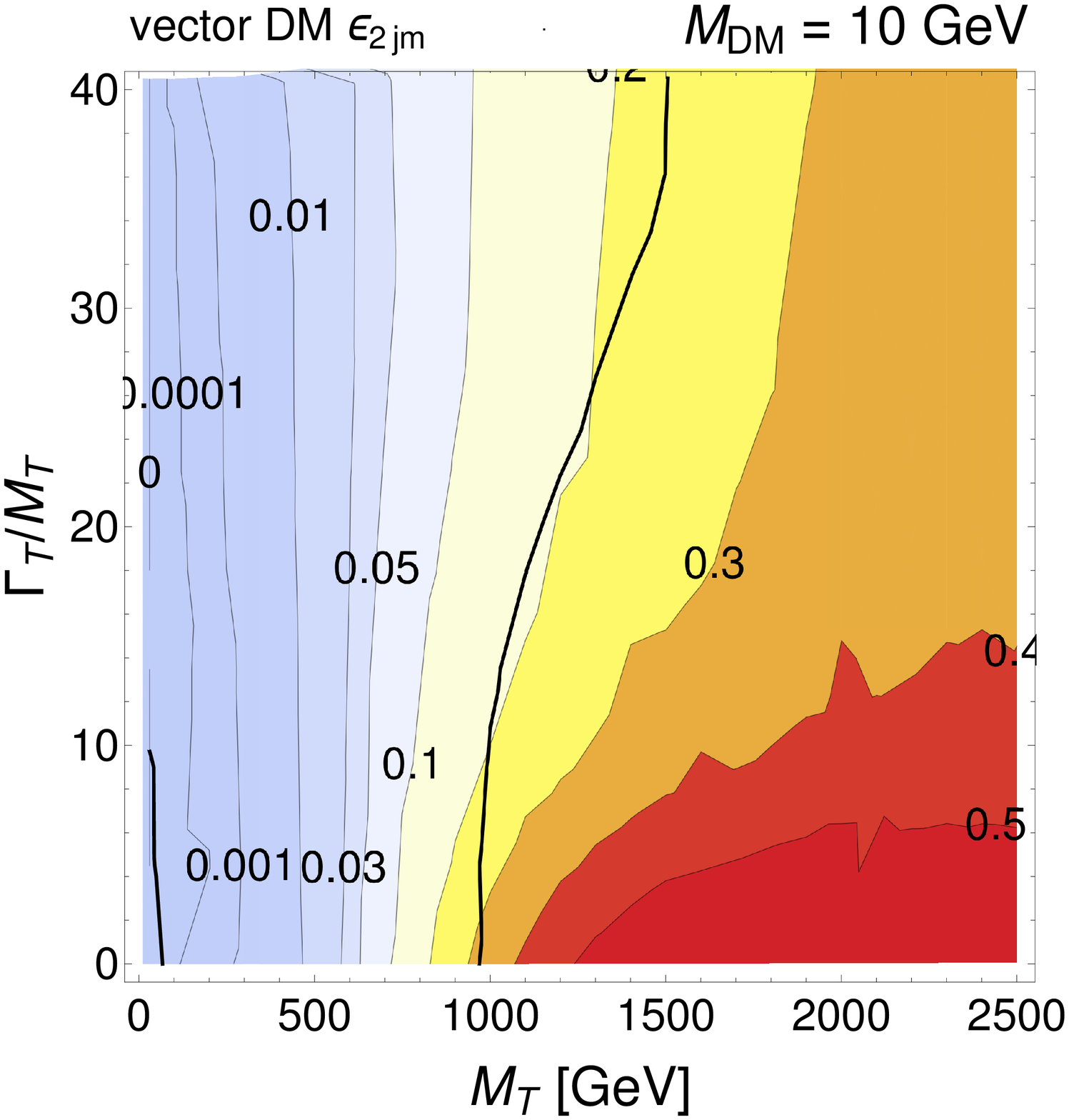}
\captionof{figure}{\label{fig:Combined1} $\sigma_S$ (left) and efficiency (right) with the exclusion line for a scalar (top row) and vector (bottom row) DM, $M_{\rm DM}$ = 10 GeV.}
\end{minipage} \hspace{2mm}
\begin{minipage}{0.44\textwidth}
We show in Fig. \ref{fig:Combined1} the value of the signal cross section and of the efficiency for one of the most important signal region for scalar and vector DM together with the exclusion line. We see that the shape of the bounds is as usual driven by the combination of cross section and efficiencies effect. Here the cross section effect is dominant: both scalar and vector DM bounds \emph{track the different behaviours in the scaling of $\sigma_S$}; the cross section scale in a different way and the bound basically follows them.

This time the hypothesis made by experimentalist are still conservative but \emph{underestimate the bounds} since the ones in the NWA are always weaker than the ones for large width which can even reach excluded values larger than $M_T = 2.5$ TeV in some cases. 
\end{minipage}

\vspace{-3mm}

\section{Exclusion limits in the $M_T-M_{DM}$ plane}
\label{sec:BellPlots}

\vspace{-1mm}

\begin{figure}
\centering
\epsfig{file=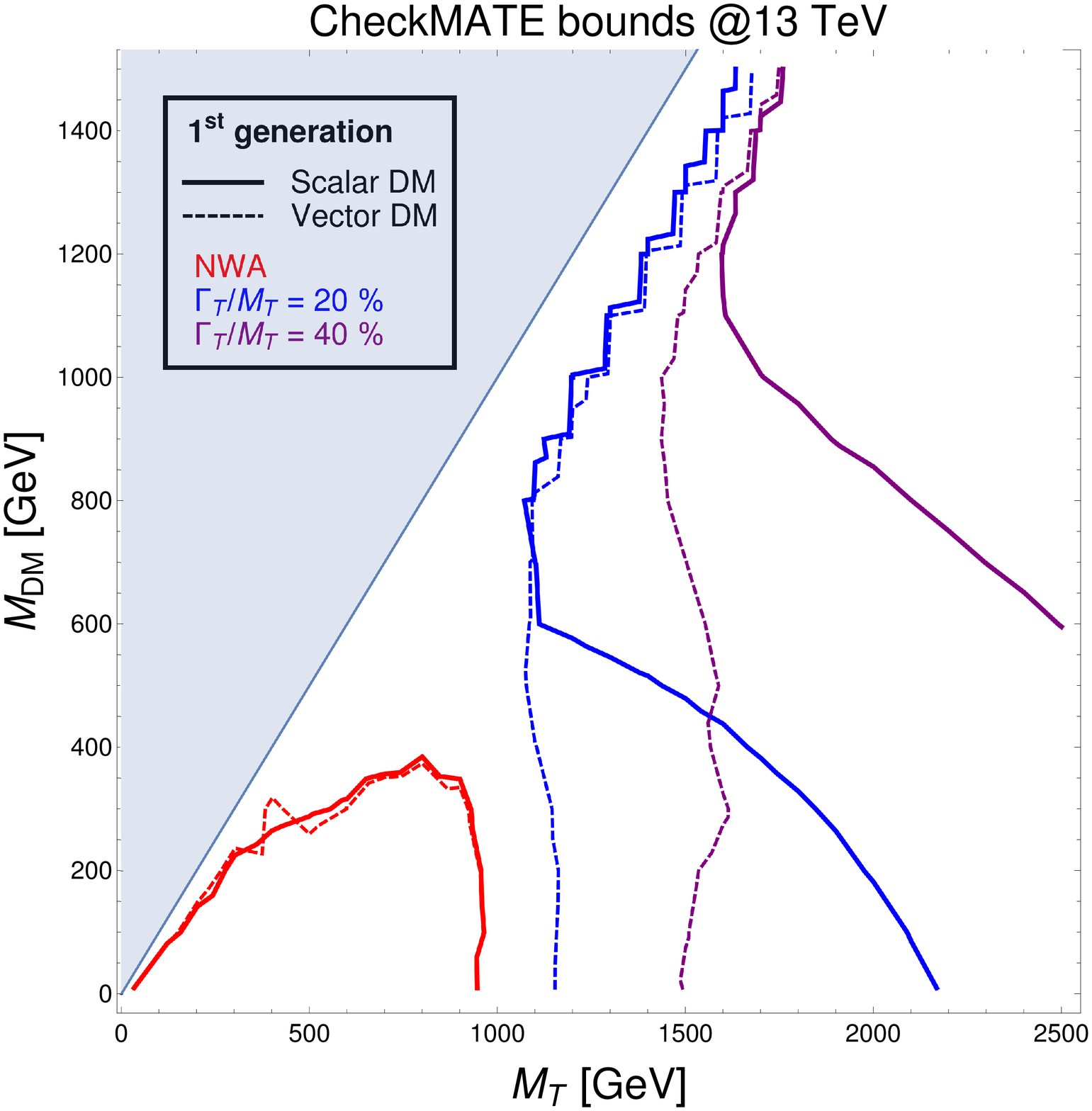, width=.32\textwidth} 
\epsfig{file=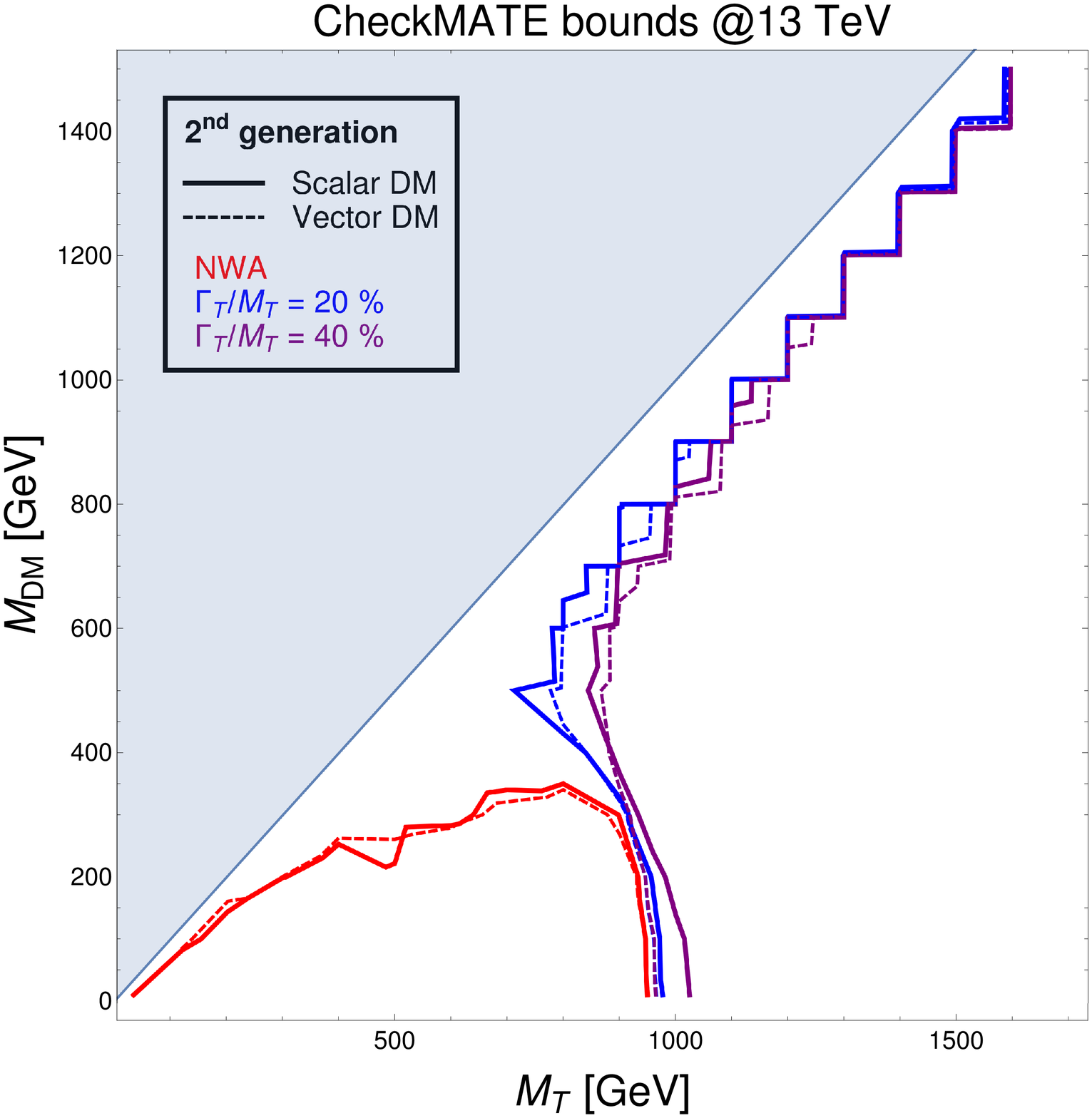, width=.32\textwidth} 
\epsfig{file=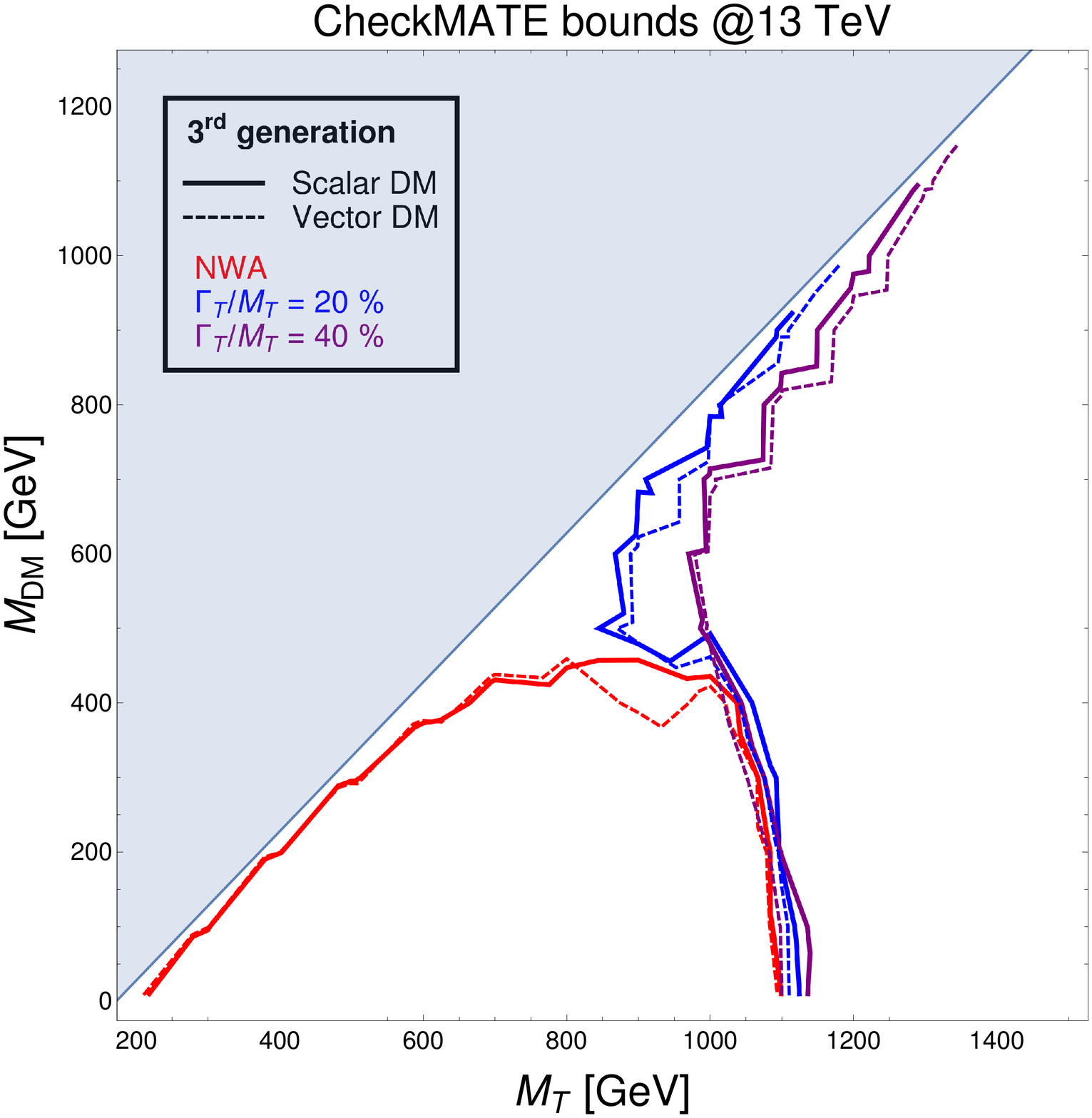, width=.32\textwidth} 
\caption{\label{fig:bellPlots} Bounds in the $(M_T, M_{\rm DM})$ plane for a coupling with first (left panel), second (center panel) and third (right panel) generations of SM quarks for different values of $\Gamma_T / M_T$.}
\end{figure}

The scenarios we are considering have three parameters: the mass of the $T$, the width of the $T$ and the mass of the DM. The exclusion bound at 2$\sigma$ will therefore identify a 3D surface in the space defined by the three parameters and therefore it is instructive to analyse the projections of this surface on the plane identified by the masses of $T$ and DM for different values of the $\Gamma_T/M_T$ ratio. Such representation is also useful to directly compare bounds on $T$ and bosonic DM with analogous results in other models, such as SUSY. Indeed the exclusion limits of SUSY searches are often presented in the $(M_{\tilde{t}}, M_{\chi_0})$ plane. We show in Fig. \ref{fig:bellPlots} the bounds in the $(M_T, M_{\rm DM})$ plane for specific values of $\Gamma_T / M_T$: the NWA case, 20\% and 40\%. We also included in this figure the results for a $T$ quark coupling to DM and the charm quark.

The qualitative behaviours of the exclusion limits strongly depend on the assumption about which SM quark generation the $T$ couples to. The bounds have the \emph{same shape like in SUSY} in the NWA while for large width the \emph{almost degenerate region is also excluded} even for large values of the masses. We also see that the \emph{width dependence} is much more important for first generation couplings, allowing us to distinguish scalar DM from vector DM in the large width regime.

To recap, the bounds obtained under the NWA are less stringent than the bounds obtained when the NWA is relaxed and the width is allowed to have large values, relative to the $T$ mass. It is remarkable, though, that different assumptions about the couplings of $T$ with different SM quark generations produce either negligible or sizably different bounds depending whether the DM is scalar or vector. This result could be exploited for the design of new experimental searches which are not only meant to discover new signals in channels with \MET\ but also to characterise these.


\section{Conclusions}
\label{sec:conclusions}

We have considered an effective model with a top-partner decaying to DM and we have studied the influence of the large width effects in the determination of the cross section and in the reinterpretation of bounds from experimental searches.

For \emph{third generation couplings}, the cross section becomes slightly larger when the width increases but this contribution is suppressed by the analysis cuts on $p_T^{\rm jet}$ and $\MET$, leading to \emph{no DM spin dependence} and only a \emph{small width dependence} on the exclusion. 
For \emph{first generation couplings}, we see a massive and width-dependent increase of the signal $\sigma_S$ due to additionnal topologies containing collinear divergences. This leads to a large width dependence of the bound. The fact that the bound follow the cross section variation which is very dependent on the DM spin leads to an \emph{important DM spin dependence} of the bounds: we are able to distinguish scalar and vector DM in this case only. 
The results for second generation fall in between the results for first and third generation.

Therefore the assumption that the NWA is a reliable approach applicable over most of the parameter space never overestimate the bound, but largely underestimate it in the case of couplings with first generation quarks. Hence, once should rescale the observed limits from established experimental analyses to the actual ones upon accounting for such effects or else attempting deploying new ones adopting different selection strategies which exalt such effects.


\bibliographystyle{JHEP}
\bibliography{XQCAT}

\end{document}